\documentclass[%
 aps,
 jmp,%
 amsmath,amssymb,
preprint,%
]{revtex4-1}
\usepackage{graphicx}
\usepackage{subfigure}
\usepackage{dcolumn}
\usepackage{bm}
\usepackage[mathlines]{lineno}
\usepackage{graphicx}
\usepackage[colorlinks,linkcolor=black,anchorcolor=black,  urlcolor=black, citecolor=blue]{hyperref}

\newcommand{\blk}{\color{black}}

\begin{document}
\title{Improved security bounds against the Trojan-Horse attack in decoy-state quantum
	key distribution}
\author{Zijian Li$^{1}$, Bingbing Zheng$^{1}$, Chengxian Zhang$^{1}$, Zhenrong Zhang$^{2}$, Hong-Bo Xie$^{3}$, Kejin Wei,$^{1,*}$}
\address{
	$^1$Guangxi Key Laboratory for Relativistic Astrophysics, School of Physical Science  and Technology,
	Guangxi University, Nanning 530004, China\\
	$^2$Guangxi Key Laboratory of Multimedia Communications and Network Technology, School of Computer, Electronics  and Information, Guangxi University, Nanning 530004, China\\	$^3$Ji Hua Laboratory, Foshan City, Guangdong Province 528200, China\\
	$^*$Corresponding author: kjwei@gxu.edu.cn}

\date{\today}

\begin{abstract}

In a quantum Trojan-horse attack (THA), eavesdroppers learn encoded information by injecting bright light into encoded or decoded devices of quantum key distribution (QKD) systems. These attacks severely compromise the security of non-isolated systems. Thus, analytical security bound was derived in previous studies. However, these studies achieved poor performance unless the devices were strongly isolated. Here, we present a numerical method for achieving improved security bound for a decoy-state QKD system under THAs. The developed method takes advantage of the well-established numerical framework and significantly outperforms previous analytical bounds regarding the achievable final key and secure transmitted distance. The results provide a new tool for investigating the efficient security bounds of THA in practical decoy-state QKD systems. This study constitutes an important step toward securing QKD with real-life components.	
\end{abstract}

\maketitle

\section{introduction}
 	Quantum key distribution (QKD)~\cite{bennett1984quantum} is a powerful tool for secure communication in the quantum information era. Currently, QKD has been experimentally proven to be applicable to several scenarios, such as optical fiber~\cite{2012wangshuang,2017Canas,2017Islam-HD,2018Yuan-fiber,2018Boaron2,2021Madi,2022huangChunfeng,2023liWei,2023grunenfelder,2023DuYongqiang,2023weiKejin} and free-space~\cite{2017Liao2,2020chenhuan,2021Avesani,2023Juboori} channels. Several small-scale networks~\cite{2011Sasaki,2019Dynes-network,2014Wang-network} have been tested in the field, and a satellite-to-ground large-scale QKD network has been reported~\cite{2021Chenyuao-network}. Recently, the recorded repeaterless distance of QKD has reached 830 km~\cite{2022Wang-TF}.
 
 Nevertheless, some significant challenges remain in the broad application of QKD technology~\cite{2016Diamanti-review,2020Xu}. In particular, bridging the gap between realistic devices and idealized models used in security proofs is crucial~\cite{2014Lo-review,2020Xu}. In particular, the eavesdropper (Eve) has exploited such deviations, such as source flaws~\cite{2015Xu,2020Pereira} and detector-efficiency mismatch~\cite{2006Makarov-mismatch,2007Qi-mismatch,2015Sajeed-mismatch,2019Wei-hacking}, to perform several subtle quantum hacking attacks~\cite{2010Lydersen-blinding,2010Xu-hacking,2015Sajeed-mismatch}.
 
 The Trojan-horse attack (THA) is a well-known hacking strategy in the QKD community~\cite{2001Vakhitov-Trojan,2006Gisin-Trojan}. In this attack, Eve illuminates bright light at encoders in legitimate users (namely, Alice and Bob) and subsequently measures the back-reflection to probe information regarding how the photon string has been encoded. In this way, Eve can break the critical assumption that no unwanted information about the settings in Alice and Bob's devices is leaked to Eve~\cite{1999Lo,2004GLLP} in most QKD security proofs without disturbing the encoded quantum states.   Such assumptions are rigorously required even in device-independent QKD~\cite{2007Acin,2021Schwonnek-DI,2022Xu-DI}. THA has been proven feasible for most practical components in QKD systems~\cite{2015Jain-Torjan,2017Sajeed-Trojan}, even in small-scale chip-based devices~\cite{2021Tan-Trojan}. The earliest versions of commercial QKD systems were reported to be of considerable risk to THA~\cite{2014Nitin-trojan-experiment}.
 
 Two major countermeasures exist for restoring QKD security during THA. The first one is the so-call ``patches," where Alice can use watch-dog devices to monitor unwanted injected light or use additional isolators to bound the intensity of the injected light. However, such countermeasures are $\emph{ad hoc}$ and can be potentially compromised by unanticipated attacks. For example, the authors~\cite{2022Ponosova} reported that the isolation component against a Trojan-horse attack could be decreased using a high-power laser. 
 
 The second countermeasure is considering the effect of side channels due to a Trojan-horse attack in the security proof. This countermeasure was first performed by Lucamarini et al.~\cite{2015Lucamarini} using a refinement of the well-known GLLP approach~\cite{2004GLLP}. In the remainder of the paper, we refer to this countermeasure the ``refined GLLP" approach for simplification. After that, in~\cite{,2016Tamaki-THA}, the asymptotic security bound for decoy-state BB84 QKD under information leakage from a legitimate user's intensity and phase device was investigated. Subsequently, the method was extended to a finite-key regime~\cite{2018Wang-THA,2022Nava-THA} and applied to decoy-state measurement-device-independent (MDI)~\cite{2021Wang-MDI-THA} and sending-or-not-sending twin-field QKD~\cite{,2021Lu-THA}. Unfortunately, the achieved secret key rate of the refined GLLP is poor unless good isolation of the transmitting unit is obtained.
 
 In this paper, we present a numerical method that improves the security bounds for decoy-state QKD protocols under THAs. Specifically, we considered two important protocols: decoy-state BB84~\cite{2005Lo,2005wangxiangbin} and decoy-state MDI-QKD~\cite{2012braunstein,2012Lo}. The decoy-state BB84 is the most mature method and is widely applied in practice. In contrast, the decoy-state MDI-QKD can remove all detector-side-channel attacks; this method has attracted considerable interest. The proposed method comprises two main components. First, we  use the numerical framework recently reported in~\cite{2021Wangwenyuan}. This allowed us to analyze the BB84 and MDI-QKD protocols using a finite number of decoy states. Second, we exploit the concept of a source-replacement scheme~\cite{2012Ferenczi-source}, which allows us to incorporate the potential information leakage due to THAs into the numerical framework. Therefore, the proposed method takes full advantage of the numerical framework for calculating key rates for practical QKD systems, which outperformed the analytical method in previous studies~\cite{2016coles-numberical,2018Winick-numberical,2021Geeorge-numberical,2020Bunandar-numberical}. Benefitting from the tight bound provided by the numerical framework, the proposed method significantly improves the achievable distance and distilled secret key rate for decoy-state BB84 and MDI-QKD protocols under THAs compared with the refined GLLP approach.  
 
 The remainder of this paper is organized as follows. In Sec.~\ref{secmethod}, the methodology used in this study is introduced. Subsequently, we use the proposed numerical method to analyze the decoy-state BB84 and MDI protocols in Sec.~\ref{Examples}. In Sec.~\ref{simulation}, we describe the simulation performed to compare the proposed method with the refined GLLP approach. Finally, we conclude this study in Sec.~\ref{Conclusion}.

\section{Method}\label{secmethod}
Here, we introduce the methodology to bound a secure key rate for decoy-state BB84 and MDI-QKD under THAs. We briefly review the numerical framework presented in~\cite{2021Wangwenyuan} and explain how to use a source-replacement scheme to incorporate potential information leakage due to THAs into the numerical framework. Furthermore, we give an intuitive explanation that our numerical method outperforms the refined GLLP method. \blk
\subsection{Numerical framework for decoy-state QKD} \label{decoy-framework}
We first provide a brief description of the numerical framework for decoy-state QKD, first introduced in~\cite{2020Li} and then developed the case of a finite number of decoy states proposed by Wang et al.~\cite{2021Wangwenyuan}. The details of this process can be found in~\cite{2021Wangwenyuan}.
A simple step-by-step descirption of the decoy-state QKD  protocol based on entanglement scheme in the numerical framework is as follows:
 \begin{enumerate}
 	\item []$\textbf{1. State preparation and measurement:}$ Alice and Bob each receive phase randomized weak coherent states in four BB84  states $\{H,V,D,A\}$ with a mean photon number $\mu$ as signal states or mean photon number $v$ as decoy states from a source. Than Alice (Bob)  randomly  select POVMs $P^{A}=\left \{P_{i}^{A}\right \}~(P^{B}=\left \{P_{j}^{B}\right \})$ and  to measure the received quantum states.
 	\item []$\textbf{2. Testing:}$   Alice and Bob select a portion of their preparation and measurement data to publish, which includes state preparation, basis selection, and measurement results for signal state events and decoy state events. This data is then used to decide whether the protocol should proceed. \blk
 	
 	\item []$\textbf{3. Announcement, sifting and postselection:}$  In each round, Bob announces the basis selection, and Alice makes her choice based on Bob's announcement. She then discards the data that is inconsistent with Bob's basis selection and informs Bob of the discarded results. This process can be represented using the Kraus operator $\left \{K_{i}\right \}$.\blk
 	
 	\item []$\textbf{4. key mapping:}$ Alice maps the data retained through the above steps to the raw key. This can be expressed using the key mapping operator $\left \{Z_{j}\right \}$.
 	\item [] $\textbf{5. Error correction and privacy amplification:}$ Alice and Bob perform standard error correction, and then they proceed to use the privacy amplification protocol to obtain the shared key.
 \end{enumerate}

 In a well-established numerical framework~\cite{2016coles-numberical}, the secure key rate calculation for the above decoy QKD was treated as an optimization problem, which can be written as follows:
\begin{equation}
R=\min _{\rho_{AB} \in S} f(\rho_{AB})-p_{\text {pass }} \times \operatorname{leak}_{\mathrm{obs}}^{\mathrm{EC}}.
\end{equation}

where $\rho_{AB}$ denotes a state shared by two remote parties, Alice and Bob; $\operatorname{leak}_{\mathrm{obs}}^{\mathrm{EC}}$ is the bits consumed during error correction; $p_{\text {pass }}$ is the probability of a signal being detected and passing the basis sifting. Moreover, $f(\rho_{AB})$ is a function related to the privacy amplification, defined as follows:

\begin{equation}
f(\rho_{AB})=D(\mathcal{G}(\rho_{AB}) \| \mathcal{Z}(\mathcal{G}(\rho_{AB}))).
\end{equation}
Here, $D(\sigma \| \tau)=\operatorname{Tr}(\sigma \log \sigma)-\operatorname{Tr}(\sigma \log \tau)$ is the relative quantum entropy. $\mathcal{G}(\rho_{AB})$ and $\mathcal{Z}(\mathcal{G}(\rho_{AB}))$ are determined by Kraus operators  $K_{i}$ (representing the measurements, public announcements and postselection process) and key map operators $Z_{j}$  (representing the key map), respectively. They satisfy the following expression:
\begin{equation}
\begin{aligned}
\mathcal{G}(\rho_{AB}) &=\sum_{i} K_{i} \rho_{AB} K_{i}^{\dagger}, \\
\mathcal{Z}(\mathcal{G}(\rho_{AB})) &=\sum_{j} Z_{j} \mathcal{G}(\rho_{AB}) Z_{j}.
\end{aligned}
\end{equation}
The density operator $\rho_{A B}$ is generally unknown. However, it can be bound by a set of states $S$ obtained from the experimental data, satisfying the following equation:
\begin{equation}
S=\left\{\rho_{AB} \in \mathbf{H}_{+} \mid \operatorname{Tr}\left(\Gamma_{k} \rho_{AB}\right)=\gamma_{k}, \forall k\right\},
\end{equation}
where $\mathbf{H}_{+}$ is the set of positive semidefinite operators, $\Gamma_{k}$ is the general positive operator-valued measures (POVM) elements representing the measurements performed by Alice and Bob, and $\gamma_{k}$ are the expectation values of the measurements. 

In particular, the study presented in~\cite{2021Wangwenyuan} has two fundamental merits to incorporate decoy-state analysis into the above numerical framework. 

First, when a phase-randomized weak coherent pulse is used in the QKD system, the photon number statistics of the pulses follow a Poisson probability distribution: 
$p_{\mu_{i}}(n)=\frac{\mu_{i}^{n}}{n !} e^{-\mu_{i}}$,  with $\mu_{i}$ being the mean intensity. In this setup, the secure key rate is generated only from single photons, which can be rewritten as
\begin{equation}
\begin{array}{l}
R\geq p_{1} \min _{\rho_{A B}^{(1)} \in S_{1}} f\left(\rho_{A B}^{(1)}\right)-p_{\text {pass }} \times \text { leak }_{\text {obs }}^{\mathrm{EC}},
\end{array}\label{keyrate}
\end{equation} 
where $p_1$ corresponds to the Poissonian distribution for sending a single photon number state, $\rho_{AB} ^{\left ( 1 \right ) } $ is the shared state conditional to a single photon being sent, and $S_1$ is the domain-bounded possible values of $\rho_{AB} ^{\left ( 1 \right ) } $.

Second, the decoy-state analysis can be used as a ``wrapper" to generate loosened bounds for $S_1$, which has the following form: 
\begin{equation}
S_{1}=\left\{\rho_{A B}^{(1)} \in \mathbf{H}_{+} \mid \gamma_{1, k}^{L} \leq \operatorname{Tr}\left(\Gamma_{k} \rho_{A B}^{(1)}\right) \leq \gamma_{1, k}^{U}, \forall k\right\},\label{bounds}
\end{equation}
where $\gamma_{1, k}^{L}( \gamma_{1, k}^{U})$ is the lower (upper) bound of the single-photon statistics obtained from the decoy-state analysis. 

Finally, the key rate for the decoy-state QKD can be calculated by running the optimization routine, based on Eq.~(\ref{keyrate}), under the constraints given in Eq.~(\ref{bounds}).
\subsection{Trojan-horse analysis} \label{Method}

In the security proof of the QKD protocol, an essential assumption is that the devices in Alice and Bob do not leak unwanted information to Eve. A simple strategy to break this assumption is the so-called Trojan-horse attack. The attack is specifically described as follows: Eve sends a bright pulse containing Trojan-horse photons to the coding devices of legitimate users. Some Trojan-horse photons are encoded with the same quantum states prepared by legitimate users and reflected back to Eve. By analyzing the reflected Trojan-horse photons, Eve can compromise the security of the QKD system.

In the security proof of the QKD protocol, an essential assumption is that the devices in Alice and Bob do not leak unwanted information to Eve. A simple strategy to break this assumption is the so-called Trojan-horse attack. The attack is specifically described as follows. Eve sends a bright pulse containing Trojan-horse photons to the coding devices of legitimate users. Some Trojan-horse photons are encoded with the same quantum states prepared by legitimate users and reflected back to Eve. By analyzing the reflected Trojan-horse photons, Eve can compromise the security of the QKD system. 

To simplify our analysis, we considered only a specific THA targeting the phase modulator in the transmitter. In this case, Eve uses a laser emitting weak coherent pulses in a coherent state $\left|\sqrt{\mu_{\text {in }}}\right\rangle$ given the average photon number $\mu_{\text {in }}$. A fraction of the pulses is encoded by carrying phase modulation information $\phi_{A}$. These pulses are then reflected back to Eve as $\left|e^{i \phi_{A}} \sqrt{\mu_{\text {out }}}\right\rangle$, where $\mu_{\text {out }}=\gamma \mu_{\text {in }}$ is the average photon number with $\gamma \ll 1$, the optical isolation of the transmitter. We can see that the light pulse retrieved by Eve is correlated to phase $\phi_{A}$, which compromises the security of the system. 

Considering a decoy-state QKD protocol, Alice sends a state $\left|\phi_{A_i}\right\rangle$ with a probability $p_i$, where $i=0...N$ and $N$ is the total number of sending quantum states. Owing to the existence of THA, the output quantum state can be written as follows:
\begin{equation}
\left|\psi_i\right\rangle=\left|\phi_{A_i}\right\rangle_{S} \otimes\left|e^{i \phi_{A_i}}\sqrt{\mu_{\text {out }}}\right\rangle_{E},\label{state}
\end{equation}
where $\phi_{A_i}$ is the specific encoded phase in the $i-$pulse; the subscript ``$S$" and ``$E$" denote the signal state prepared by Alice and the Trojan-horse state obtained by Eve, respectively. Here, we assume that the THA does not affect the decoy-state analysis. Hence, the state can be straightforward in the single-photon form following the analysis described in~\cite{2015Lucamarini,2021Wangwenyuan}. 

To incorporate the Trojan-horse analysis into the decoy-state numerical framework, we use a source-replacement method. The method has been used to recast prepare-and-measure protocols as entanglement-based protocols~\cite{2009Scarani-review}. The encoder can recast the state in Eq. (\ref{state}) as the entanglement state using the source-replacement scheme as follows: 
\begin{equation}
|\Phi\rangle_{{AA}^{\prime}E}=\sum_{{i}} \sqrt{p_{i}}|{i}\rangle_{A}\left|\psi_i\right\rangle_{A^{\prime}E},
\end{equation}
where $A$ is a registration system for storing the information $|{i}\rangle_A$ regarding which state Alice has prepared. Subsequently, Alice keeps system $A$ and sends system $A'$ to Bob through a quantum channel $\xi$, so that the final joint state is as follows:
\begin{equation}
\rho_{A B }=\left(\mathbb{I}_{A} \otimes \xi\right)\operatorname{Tr}_{E}\left(|\Phi\rangle_{A A'E} \langle\Phi|\right)\label{rho},
\end{equation}
where $\mathbb{I} _{A} $ denotes the identity channel on A. By applying the numerical optimization method described in Sec.~\ref{decoy-framework} to state $\rho_{A B}$ in Eq.~(\ref{rho}), we can calculate the final key rate by considering THAs. Furthermore, we must add additional constraints to account for the particular form of $\rho_{A B}$.

For the measurement, we have the following constraints:
\begin{equation}\label{yueshu}
\operatorname{tr}\left(\left(\mathrm{P}_{j}^{\mathrm{A}} \otimes \mathrm{P}_{i}^{\mathrm{B}}\right)	\rho_{A B}\right)=p_{j i}.
\end{equation}

 In addition to measurement constraints, in order to optimize the key rate in the presence of THA, the assumption in~\cite{2016coles-numberical} is used: Alice has well characterized her source. In other words, Alice strictly knows the following states
\begin{equation}
\rho_{A}=\operatorname{Tr}_{B}\left(\rho_{AB}\right)=\operatorname{Tr}_{\mathrm{A}^{\prime}E}\left(|\Phi\rangle_{AA^{\prime}E}\langle\Phi|\right).
\end{equation}
Therefore, we need to add constraints of the form 
\begin{equation}
\operatorname{Tr}\left(\left(\Theta_{j} \otimes \mathbb{I}_{B}\right) \rho_{A B}\right)=\theta_{j},
\end{equation}
into the constraints Alice and Bob have on their states, where $\{\Theta_{j}\}$ is a set of the tomographic observables on system $A$.

 We can see that the constraint of Eq.~(\ref{yueshu}) is the measured value of the single photon component. In the case of WCP source, we cannot know it. However, we can use the decoy state analysis technology to obtain its estimated value, so we will have the following constraints:

\begin{equation}
\gamma_{1, k}^{L} \leq \operatorname{tr}\left(\left(\mathrm{P}_{j}^{\mathrm{A}} \otimes \mathrm{P}_{i}^{\mathrm{B}}\right)	\rho_{A B}\right) \leq \gamma_{1, k}^{U}\label{bounds},
\end{equation}
where $\gamma_{1, k}^{L}$ and $\gamma_{1, k}^{U}$ are the  lower and upper bounds of POVM measurement $\mathrm{P}_{j}^{\mathrm{A}} \otimes \mathrm{P}_{i}^{\mathrm{B}}$ respectively. For the above constraints, we need corresponding statistical data. In practice, we can get these statistics from experiments. In the simulation, we can obtain the original statistical information of these WCP sources through the  channel model. See the appendix \ref{Channelmodel} for the specific channel model.

\subsection{Improved security bound under THA attack}\label{improvereson}
In this section, we will explain in detail why our numerical method enables improving the key rate under THA attacks. 

For clarity, let's  briefly review the refined GLLP's method.
In refined-GLLP
approach presented in~\cite{2015Lucamarini},  the final key rate in presence of THA can be expressed as  
\begin{equation}
	R=1-h_{2}(e^{'}_{X})-h_{2}(e_{Z}).
\end{equation}
Here, $e_{Z}$ is the quantum bit error rate of the signal states in the $Z$ basis,  $e_{X}^{\prime}$ is phase error rate of a single photon under THA, $h_{2} \left ( x \right )$ is the binary Shannon information function. Finally, based on the analysis in~\cite{2015Lucamarini}, $e_{X}^{\prime}$ is given by ``Bloch Sphere bound"~\cite{2006LP}, which can be expressed the following equation:
\begin{equation}
	\begin{aligned}
		e_{X}^{\prime}=& e_{X}+4 \Delta^{\prime}\left(1-\Delta^{\prime}\right)\left(1-2 e_{X}\right) \\
		&+4\left(1-2 \Delta^{\prime}\right) \sqrt{\Delta^{\prime}\left(1-\Delta^{\prime}\right) e_{X}\left(1-e_{X}\right)}, \\
		\Delta^{\prime}=& \frac{\Delta}{{Y}}, \\
		\Delta=& \frac{1}{2}\left[1-\exp \left(-\mu_{\text {out }}\right) \cos \left(\mu_{\text {out }}\right)\right],
	\end{aligned}
\end{equation}
where $e_{X}$ is the quantum bit error rate in the $X$ basis, and $Y$ is defined as ${Y}:=\min \left[{Y}_{Z}^1, {Y}_{X}^1\right]$ with ${Y}_{Z}^1$ (${Y}_{X}^1$) as the single-photon yields on the $Z$ ($X$) basis.

\begin{figure}[!ht]
	\centering
	\includegraphics[width=\linewidth]{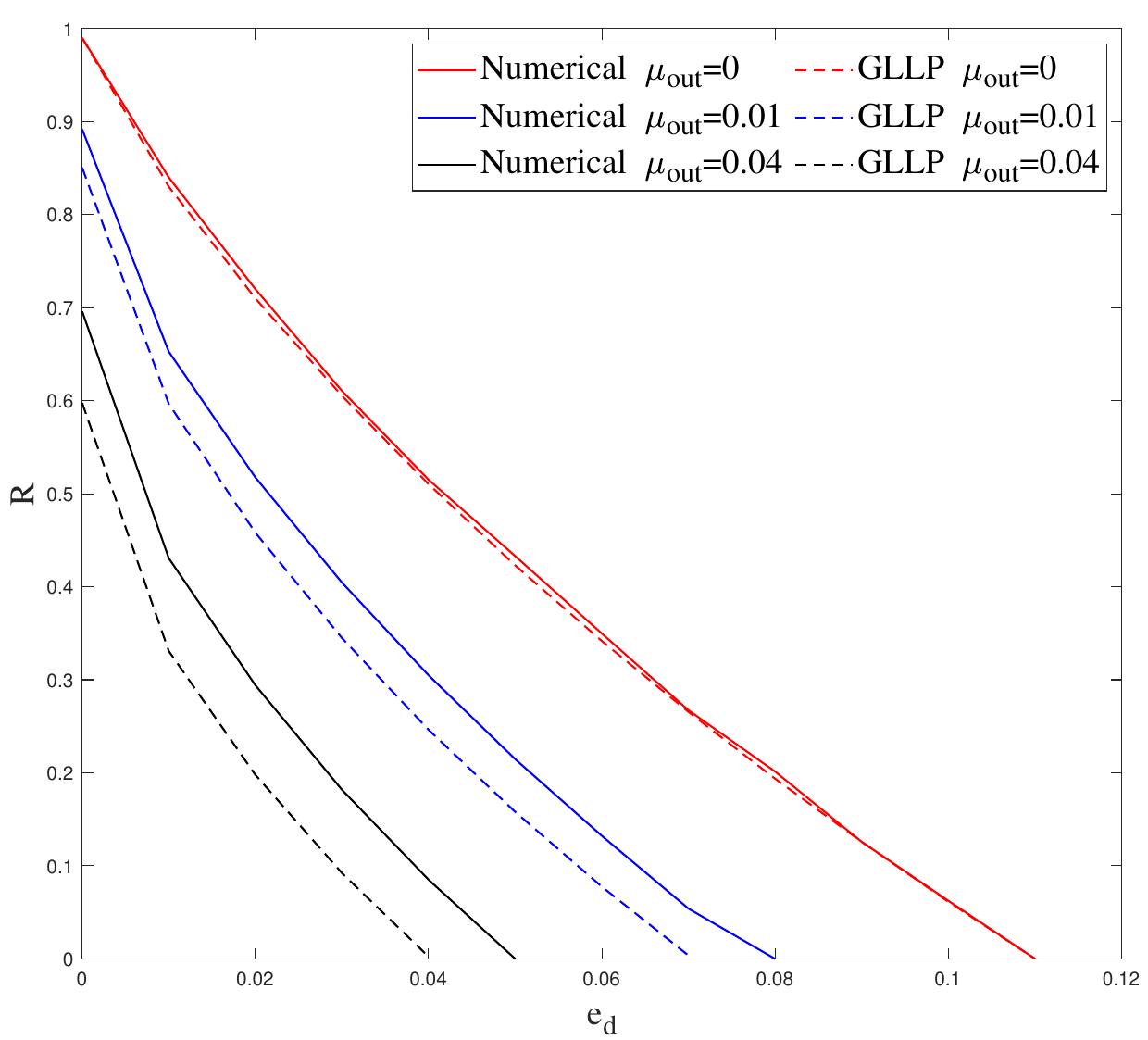}
	\caption{Simulation result of key rate vs error rate for single photon BB84 protocol under Trojan attack. The key rate is plotted for different values of $\mu_{out}$. $e_d$ represents the intrinsic misalignment.}\label{sigle}
\end{figure}

From the above description, we can see that the refined GLLP's method essentially bounds the Eve's information by using several inequalities, which have looseness bound, resulting a poor performance of the final key rate. In contrast, the numerical method presented in Ref. ~\cite{2018Winick-numberical} provides a tight bound by using two-step optimization. This advantages are feasible even in presence of THA. For a first shot, we take an ideal case (single photon case, no loss, no dark counting) as an example. The results are shown in the Fig.~\ref{sigle}. It can seen that the numerical method outperfoms the refined GLLP's method with different leaked intensity $\mu_{out}$. We remark that these results also are proved in Ref.~\cite{2018Winick-numberical} but it did not consider decoy-state analysis.

\section{Examples}\label{Examples}
This section provides examples of the proposed numerical approach applied to specific protocols, including decoy-state BB84 and MDI under THAs.

\subsection{Decoy-state BB84 QKD}
Here, we consider the phase-encoding BB84 scheme. Typical units for generating the four BB84 quantum states can be found in the asymmetrical, fiber-based, phase-modulated QKD setup in Ref.~\cite{2018Yuan-fiber}. In this setup, we can write the $Z$-basis states corresponding the phase $\phi_{A}=\{0,\pi\}$ as $\left|z_{\pm}\right\rangle:=\frac{1}{\sqrt{2}}\left(|1\rangle_{L}|0\rangle_{M} \pm|0\rangle_{L}|1\rangle_{M}\right)$. Moreover, the $X$-basis states corresponding the phase $\phi_{A}=\{\pi/2,3\pi/2\}$ can be written as $\left|x_{\pm}\right\rangle:=\frac{1}{\sqrt{2}}\left(|1\rangle_{L}|0\rangle_{M} \pm i|0\rangle_{L}|1\rangle_{M}\right)$ for $X$ basis, where $\left(|n\rangle_{L}\right)$ and $\left|n\rangle_{M}\right)$ denote the $n-$photon passing the long and short arm of interferometer, respectively. In the presence of THAs, the output quantum state from Alice can be rewritten as follows:
\begin{equation}
\begin{aligned}
\left|\psi_{z+}\right\rangle=\left|z_{+}\right\rangle_{S} \otimes\left|+\sqrt{\mu_{\text {out }}}\right\rangle_{E},\\ 	
\left|\psi_{z-}\right\rangle=\left|z_{-}\right\rangle_{S} \otimes\left|-\sqrt{\mu_{\text {out }}}\right\rangle_{E},\\ 
\left|\psi_{x+}\right\rangle=\left|x_{+}\right\rangle_{S} \otimes\left|+i \sqrt{\mu_{\text {out }}}\right\rangle_{E},\\
\left|\psi_{x-}\right\rangle=\left|x_{-}\right\rangle_{S} \otimes\left|-i \sqrt{\mu_{\text {out }}}\right\rangle_{E}.
\end{aligned}\label{bb84-states}
\end{equation}

By following the method described in Sec.~\ref{Method}, Alice can be regarded as preparing the following states:
\begin{equation}
\begin{aligned}
|\Phi\rangle_{AA^{\prime}E}=\sqrt{{p_{\mathrm{z+}}}}|0\rangle_{A}\left|\psi_{\mathrm{z}+}\right\rangle_{A'E}+\sqrt{{p_{\mathrm{z-}}}}|1\rangle_{A}\left|\psi_{\mathrm{z}-}\right\rangle_{A'E}+\\\sqrt{{p_{\mathrm{x+}}}}|2\rangle_{A}\left|\psi_{\mathrm{x}+}\right\rangle_{A'E}+\sqrt{{p_{\mathrm{x-}}}}|3\rangle_{A}\left|\psi_{\mathrm{x}-}\right\rangle_{A'E},
\end{aligned}
\end{equation}
where Alice, with a probability $p_{\alpha}$, is the probability of sending state $\alpha$ $\in$ $\{z+,  z-, x+,  x-\}$ from Alice.
In the communication stage, Alice first sends part of the state $|\Phi\rangle_{A A^{\prime}}$ to Bob (i.e., system $A'$) using the quantum channel $\xi$; thus, we obtain the following joint state:
\begin{equation}
\rho _{A B}=\left(\mathbb{I}_{A} \otimes \xi\right)\operatorname{Tr}_{E}\left(|\Phi\rangle_{AA^{\prime}E}(\Phi \mid)\right..
\end{equation}
The measurement considered the following constraints (see Appendix~\ref{BB84measurementoperator} for detailed description of measurement operators):
\begin{equation}
\begin{aligned}
\operatorname{Tr}\left(\left(\mathrm{P}_{j}^{\mathrm{A}} \otimes \mathrm{P}_{i}^{\mathrm{B}}\right)	\rho_{A B}\right)=p_{j i}.\\
\operatorname{Tr}\left(\left(\Theta_{j} \otimes \mathbb{I}_{B}\right) \rho_{A B}\right)=\theta_{j}.	
\end{aligned}
\end{equation}
where $\Theta_{j}$ is the tomographic operator of system $A$. 

In this study, we compare our method to the refined-GLLP
approach presented in~\cite{2015Lucamarini}, where the final key rate is 
\begin{equation}
\left.R=p_{Z}^{2} p_{1} Y_{1}\left[1-h_{2}\left(e_{X}^{\prime} \right)\right]-p_{Z}^{2} Q_{\mu} fh_{2}\left(E_{\mu}\right)\right).
\end{equation}
Here, $Q_{\mu}$ is the gain of the signal states, $E_{\mu}$ is the quantum bit error rate of the signal states, $p_{1}$ is the probability of sending a single photon, $Y_{1}$ and $e_{X}^{\prime}$ are the yield and error rate of a single photon under THA, estimated following the decoy-state analysis. Moreover, $f$ is the error correction efficiency. Finally, based on the analysis in~\cite{2015Lucamarini}, $e_{X}^{\prime}$ is given by the following equation:
\begin{equation}
\begin{aligned}
e_{X}^{\prime}=& e_{X}+4 \Delta^{\prime}\left(1-\Delta^{\prime}\right)\left(1-2 e_{X}\right) \\
&+4\left(1-2 \Delta^{\prime}\right) \sqrt{\Delta^{\prime}\left(1-\Delta^{\prime}\right) e_{X}\left(1-e_{X}\right)}, \\
\Delta^{\prime}=& \frac{\Delta}{{Y}}, \\
\Delta=& \frac{1}{2}\left[1-\exp \left(-\mu_{\text {out }}\right) \cos \left(\mu_{\text {out }}\right)\right],
\end{aligned}
\end{equation}
where $e_{X}$ is the quantum bit error rate in the $X$ basis, and $Y$ is defined as ${Y}:=\min \left[{Y}_{Z}^1, {Y}_{X}^1\right]$ with ${Y}_{Z}^1$ (${Y}_{X}^1$) as the single-photon yields on the $Z$ ($X$) basis.

\subsection{Decoy-state MDI QKD}
In the phase-encoding MDI protocol, Alice and Bob both prepare four BB84 states using similar setups of the BB84 QKD system; an example is shown in~\cite{2019Liuhui}. Therefore, the output states sent from Alice and Bob have the same form as in Eq.~(\ref{bb84-states}).

Similarly, using the source-replacement scheme, Alice prepares the entangled states as follows:

\begin{equation}
\begin{aligned}
|\Phi\rangle_{AA^{\prime}E_{A}}=\sqrt{{p_{\mathrm{z+}}}}|0\rangle_{A}\left|\psi_{\mathrm{z}+}\right\rangle_{A'E_{A}}+\sqrt{{p_{\mathrm{z-}}}}|1\rangle_{A}\left|\psi_{\mathrm{z}-}\right\rangle_{A'E_{A}}+\\\sqrt{{p_{\mathrm{x+}}}}|2\rangle_{A}\left|\psi_{\mathrm{x}+}\right\rangle_{A'E_{A}}+\sqrt{{p_{\mathrm{x-}}}}|3\rangle_{A}\left|\psi_{\mathrm{x}-}\right\rangle_{A'E_{A}}.
\end{aligned}
\end{equation}
Moreover, Bob prepares the entangled states as follows:
\begin{equation}
\begin{aligned}
|\Phi\rangle_{BB^{\prime}E_{B}}=\sqrt{{p_{\mathrm{z+}}}}|0\rangle_{A}\left|\psi_{\mathrm{z}+}\right\rangle_{A'E_{B}}+\sqrt{{p_{\mathrm{z-}}}}|1\rangle_{A}\left|\psi_{\mathrm{z}-}\right\rangle_{A'E_{B}}+\\\sqrt{{p_{\mathrm{x+}}}}|2\rangle_{A}\left|\psi_{\mathrm{x}+}\right\rangle_{A'E_{B}}+\sqrt{{p_{\mathrm{x-}}}}|3\rangle_{A}\left|\psi_{\mathrm{x}-}\right\rangle_{A'E_{B}}.
\end{aligned}
\end{equation}
Here, $E_A(E_B)$ denotes the system obtained by Eve in the channel between Alice (Bob) and Charlie. At the communication stage, Alice and Bob first send part of the state $|\Phi\rangle_{A A^{\prime}}$ and $|\Phi\rangle_{B B^{\prime}}$ to Charlie (i.e., states in system $A^{\prime}$ and system $B^{\prime}$) through the quantum channel $\xi$ to obtain the final joint state
\begin{equation}
\begin{aligned}
\rho_{A B C}=\left(\mathbb{I}_{A} \otimes \xi\right) \operatorname{Tr}_{E_{A}}\left(|\Phi\rangle_{AA^{\prime}E_{A}}\langle\Phi|\right) \otimes\\\left(\mathbb{I}_{B} \otimes \xi\right)
\operatorname{Tr}_{E_{B}}\left(|\Phi\rangle_{BB^{\prime}E_{B}}\langle\Phi|\right).
\end{aligned}
\end{equation}

The measurement considered the following constraints (See Appendix~\ref{BB84measurementoperator} for detailed description of measurement operators):
\begin{equation}
\operatorname{tr}\left(\left(\mathrm{P}_{j}^{\mathrm{A}} \otimes \mathrm{P}_{i}^{\mathrm{B}} \otimes \mathrm{P}_{{k}}^{\mathrm{C}}\right) \rho_{A B C})=\mathrm{p}_{j ik}\right.,
\end{equation}

In this scenario, because system A and B has well characterized source, we have the following expression:
\begin{equation}
\begin{aligned}
\rho_{A B}=\operatorname{Tr}_{\mathrm{C}}\left(\rho_{\mathrm{ABC}}\right)=\operatorname{Tr}_{A'E_{A}}\left(|\Phi\rangle_{AA^{\prime}E_{A}}\langle\Phi|\right)\otimes\\\operatorname{Tr}_{B'E_{B}}\left(| \Phi\rangle_{BB^{\prime}E_{B}}\langle\Phi|\right).
\end{aligned}
\end{equation}
Therefore, the following constraints must be added:
\begin{equation}
\operatorname{Tr}\left(\left(\Theta_{{j}}^{\mathrm{A}} \otimes \Theta_{{i}}^{\mathrm{B}} \otimes \mathbb{I}_{\mathrm{C}}\right) \rho_{A B C}\right)=\theta_{{ji}},
\end{equation}
where $\Theta_{\mathrm{j}}^{\mathrm{A}}$ and $\Theta_{\mathrm{K}}^{\mathrm{B}}$ 
are the tomographic operators in space A and space B, respectively.

Similarly, we compared the proposed approach with the refined GLLP for MDI-QKD presented in~\cite{2021Tan-Trojan}. The key rate is as follows:
\begin{equation}
R=p_{Z}^{2} p_{11} Y_{11}^{Z}\left[1-h_{2}\left(e_{11}^{X'}\right)\right]-p_{Z}^{2} Q_{\mu \mu}^{Z} fh_{2}\left(E_{\mu \mu}^{Z}\right),
\end{equation}
where $Q_{\mu \mu }^{Z} $is the signal state gain, $E_{\mu \mu }^{Z} $ is the signal state quantum bit error rate, $p_{11}$ is the probability that Alice and Bob simultaneously send a single photon, $f$ is the error correction efficiency, and $Y_{11}^{Z}$ is the single-photon yield in the $Z$ basis, directly estimated by the decoy-state analysis). Moreover, $e_{11 }^{X}$ is the single-photon error rate in the $X$ basis and satisfies the following equation:
\begin{equation}
\begin{aligned}
e_{11}^{X'}=& e_{11, \mathrm{bit}}^{X}+4 \Delta^{\prime}\left(1-\Delta^{\prime}\right)\left(1-2 e_{11}^{X, \mathrm{bit}}\right) \\
&+4\left(1-2 \Delta^{\prime}\right) \sqrt{\Delta^{\prime}\left(1-\Delta^{\prime}\right) e_{11, \mathrm{bit}}^{X}\left(1-e_{11, \mathrm{bit}}^{X}\right)},
\end{aligned}
\end{equation}
where $e_{11, \mathrm{bit}}^{X}$ is the single-photon error rate estimated using the decoy-state analysis.
Moreover, $e_{11}^{X}$ is the quantum bit error rate when THA does not exist, and $ \Delta^{\prime}$ satisfies
\begin{equation}	
\Delta^{\prime}=\frac{\Delta}{Y_{11}^{Z}},
\end{equation}
where
\begin{equation}	
\Delta=\frac{1}{2}\left[1-\exp \left[-(\mu _{out}^{Alice}+\mu _{out}^{Bob}) \right] \cos \left[ \frac{1}{2}(\mu _{out}^{Alice}+\mu _{out}^{Bob}) \right]^{2}\right],
\end{equation}
where $\mu _{out}^{Alice}$ and $ \mu _{out}^{Bob}$ represent the intensities reflected from Alice and Bob, respectively.

\blk
\section{Simulation}\label{simulation}
 In this section, we first present a simulation performed to compare the proposed method with the refined GLLP approach in~\cite{2015Lucamarini}.  The parameters used in our simulation were extracted from previous related work. The parameters are summarized in Tab.~\ref{Pa}. The specific channel models for simulating the raw statistics obtained from WCP sources for BB84 and MDI-QKD are shown in~Appendix~\ref{Channelmodel}.  

\begin{table}[h] 
	\setlength\tabcolsep{3pt} 
	\centering
	\caption{Parameters for the experiments and numerical simulations. $\eta_{d}$ is the detection efficiency, $e_d$ denotes the optical misalignment, $Y_0$ is the dark count rate, and $f$ is the error correction efficiency.}
	\begin{tabular}{cccccccc}
		\hline\hline
		& Parameter&Reference&Protocol&$e_d$&$Y_{0}$&$\eta_{d}$&$f$\\ \hline
		&Case 1&Ref.~\cite{2015Lucamarini}&BB84&0.01&$1\times10^{-5}$&0.125&1.2 \\ \hline
		&Case 2&Ref.~\cite{2021Tan-Trojan}&MDI&0.02&$8\times10^{-8}$&0.495&1.16 \\ \hline
	\end{tabular}\label{Pa}
\end{table}	
\begin{figure}[ht]
	\centering
	\includegraphics[width=\linewidth]{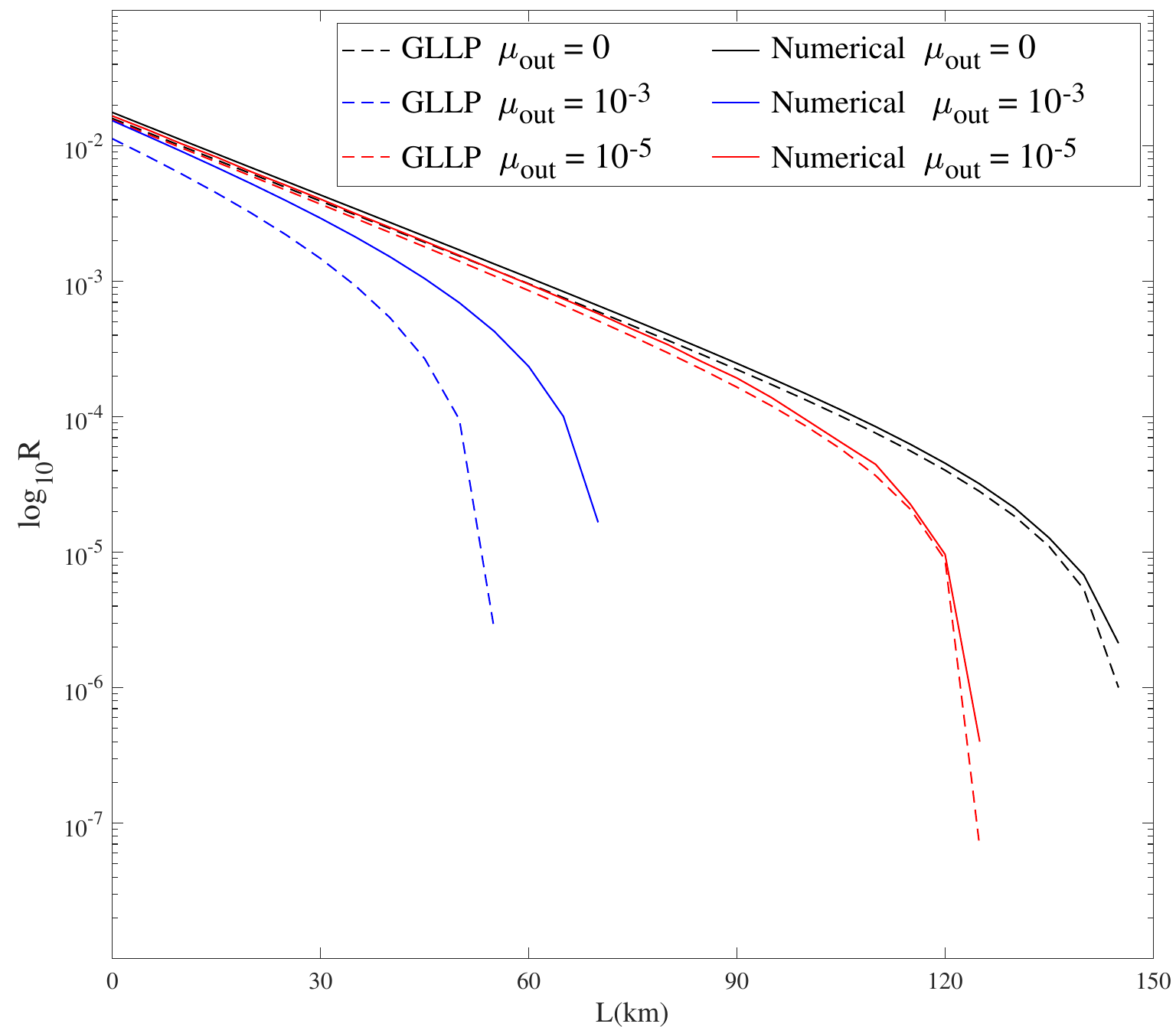}
	\caption{Simulation results for decoy-state BB84 under THA. $\mu_{\text {out }}$ is the average number of photons reflected by THA. The black, blue, and red solid (dotted) lines are the key rate of using (our numerical method) the refined GLLP approach for various values of the parameters $\mu_{\text {out }}$. For each point, we optimized the intensity of signal states and set the intensity of decoy state $\nu_{1}=0.02$ and the intensity of vacuum state $\nu_{2}=0.001$.}.\label{BB84rate}
	
\end{figure}
\begin{figure}[!ht]
	\centering
	\includegraphics[width=\linewidth]{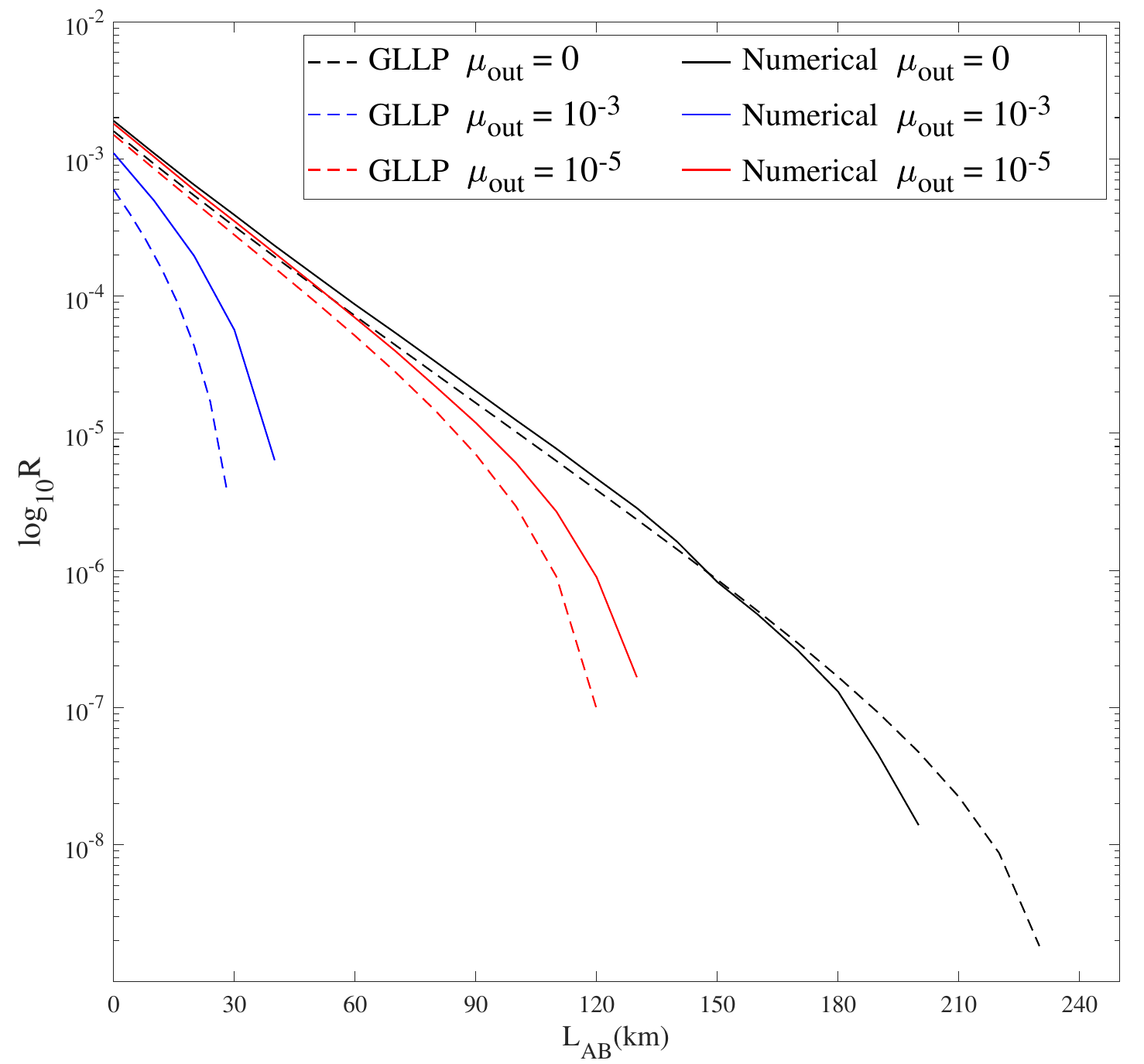}
	\caption{Simulation results for decoy-state MDI under THA. $\mu_{\text {out }}$ is the average number of photons reflected by THA. The black, blue, and red solid (dotted) lines are the key rate of using (our numerical method) the refined GLLP approach for various values of the parameters $\mu_{\text {out }}$. For each point, we optimized the intensity of signal states and set the intensity of decoy state $\nu_{1}=0.02$ and the intensity of vacuum state $\nu_{2}=0.001$.}\label{MDIrate}
\end{figure}

In Fig.~\ref{BB84rate}, we plot the key rate of the decoy-state BB84 protocol using our method and compare it with the result obtained using the GLLP approach for different reflected Trojan-horse intensity $\mu_{\text {out }}$ values. The parameters used were the same as those used in a previous study analyzing decoy-state BB84 using a refined GLLP approach. The specific values are listed for Case 1  in Tab.~\ref{Pa}. As shown in Fig.~\ref{BB84rate}, the proposed numerical method provides a higher key rate and longer transmission distance than the refined GLLP approach. In particular, the advantage of our method is evident at a large value of  $\mu_{\text {out }}=10^{-3}$. At this large value, the refined GLLP approach can only reach a secure distance of  55 km, which is only $78.6\%$ of that obtained using the proposed numerical approach.

In Fig.~\ref{MDIrate}, we depict the key rate for MDI-QKD using the proposed method. The parameters used were extracted from Refs.~\cite{2021Tan-Trojan} and are listed as Case 2 in Tab.~\ref{Pa}. Fig.~\ref{MDIrate} shows a behavior similar to that of decoy-state BB84. In other words, our numerical method outperforms the refined GLLP approach, providing a higher key rate and allowing a longer secure distance. In particular, when $\mu_{\text {out }}=10^{-3}$, the proposed numerical method can provide a communication distance of 40 km. In contrast,  the transmitted distance is 28 km using the refined GLLP approach. It should be noted that when $\mu_{\text {out }}=0$, long distance, the existence of constraint noise makes the optimization very difficult and leads to incomplete optimization, so the key rate is lower than that of GLLP approach. However, this is not caused by numerical method, but only due to technical problems.

\section{Conclusion}\label{Conclusion}

In summary, we developed a numerical method for calculating the security bounds for decoy-state QKD protocols under THAs. Benefitting from the tight security bound provided by the numerical framework, the proposed method improved the achieved secure key rate and prolonged the maximum communication distance for the decoy-state BB84 QKD and MDI-QKD protocols under THAs. In future research, the numerical method could be extended to more general Trojan-horse attacks, as described in~\cite{2016Tamaki-THA}, or jointly incorporating more imperfections into the numerical framework, as in the GLLP-based analysis in~\cite{2021Sun}. Furthermore, this study only considered the asymptotic case. Thus, its combination with a finite-key analysis~\cite{,2020Bunandar-numberical,2021Geeorge-numberical} could be the subject of future work. 

\section{Acknowledgments}
We thank Shi-Hai Sun and Wenyuan Wang for their helpful discussions. This study was supported by the National Natural Science Foundation of China (Nos. 62171144 and 62031024 and 11905065), the Guangxi Science Foundation (Nos.
2021GXNSFAA220011 and 2021AC19384), and the Open Fund of IPOC (BUPT) (No. IPOC2021A02).
\section{Data availability}
Data will be made available on reasonable request 
\bibliography{Trojan}
\appendix
\section{measurement operator, Kraus operator, key mapping} \label{BB84measurementoperator}
In this section, we will describe the specific forms of operators required in BB84 and MDI protocols. Our protocol description model is similar to Ref.~\cite{2021Wangwenyuan}. 
\subsection{BB84}
The measurement operator is:

	\begin{tabular}{cccccccc}
		\hline
		&$P_{i}^{A}$&&$|0\rangle\langle 0|$ &$|1\rangle\langle 1|$&$|2\rangle\langle 2|$&$|3\rangle\langle 3|$&\\ \hline
		&$P_{i}^{B}$&&$|Z_{+} \rangle\langle Z_{+}| \oplus 0$&$|Z_{-} \rangle\langle Z_{-}|\oplus 0$ &$|X_{+} \rangle\langle X_{+}|\oplus 0 $&$|X_{-} \rangle\langle X_{-}|\oplus 0$&$ 1-\sum_{i=1}^{4} P_{i}^{B}$\\ \hline
	\end{tabular}

The tomographic scanning operator is:
\begin{equation}
	|i\rangle\left\langle\left. j\right|_{A} \otimes \mathbb{I}_{\text {dim }_{B}}.\right.
\end{equation}
The Kraus operator is:
\begin{equation}
		\begin{array}{l}
			K_{Z}=\left[\left(\begin{array}{l}
				1 \\
				0
			\end{array}\right) \otimes\left(\begin{array}{llll}
				1 & & & \\
				& 0 & & \\
				& & 0 & \\
				& & & 0
			\end{array}\right)+\left(\begin{array}{l}
				0 \\
				1
			\end{array}\right) \otimes\left(\begin{array}{llll}
				0 & & & \\
				& 1 & & \\
				& & 0 & \\
				& & & 0
			\end{array}\right)\right] \\
			\sqrt{p_{Z}}\left(\begin{array}{lll}
				0 & & \\
				& 1 & \\
				& & 1
			\end{array}\right) \otimes\left(\begin{array}{l}
				1 \\
				0
			\end{array}\right) ,\\
			K_{X}=\left[\left(\begin{array}{l}
				1 \\
				0
			\end{array}\right) \otimes\left(\begin{array}{llll}
				0 & & & \\
				& 0 & & \\
				& & 1 & \\
				& & & 0
			\end{array}\right)+\left(\begin{array}{l}
				0 \\
				1
			\end{array}\right) \otimes\left(\begin{array}{llll}
				0 & & & \\
				& 0 & & \\
				& & 0 & \\
				& & & 1
			\end{array}\right)\right] \\
			\sqrt{p_{X}}\left(\begin{array}{ccc}
				0 & & \\
				& 1 & \\
				& & 1
			\end{array}\right) \otimes\left(\begin{array}{l}
				0 \\
				1
			\end{array}\right). \\
		\end{array}
\end{equation}
 While the key maps are:\blk
\begin{equation}
\begin{aligned}
	Z_{1} &=\left(\begin{array}{ll}
		1 & 0 \\
		0 & 0
	\end{array}\right) \otimes \mathbb{I}_{\operatorname{dim}_{A} \times \operatorname{dim}_{B} \times 2}, \\
	Z_{2} &=\left(\begin{array}{ll}
		0 & 0 \\
		0 & 1
	\end{array}\right) \otimes \mathbb{I}_{\operatorname{dim}_{A} \times \operatorname{dim}_{B} \times 2}.
\end{aligned}
\end{equation}
\subsection{MDI}\label{MDImeasurementoperator}
The measurement operator is

\begin{tabular}{cccccccc}
	\hline
	&$P_{i}^{A}$&&$|0\rangle\langle 0|$ &$|1\rangle\langle 1|$&$|2\rangle\langle 2|$&$|3\rangle\langle 3|$\\ \hline
	&$P_{i}^{B}$&&$|0\rangle\langle 0|$ &$|1\rangle\langle 1|$&$|2\rangle\langle 2|$&$|3\rangle\langle 3|$\\ \hline
		&$P_{i}^{C}$&&$ \left|\Phi^{+}\right\rangle_{a b}\left\langle\left.\Phi^{+}\right|_{a b}\right.$ &$ \left|\Phi^{+}\right\rangle_{a b}\left\langle\left.\Phi^{+}\right|_{a b}\right.$&$1-\sum_{i=1}^{2} P_{i}^{C}$\\ \hline
\end{tabular}

The tomographic scanning operator is
\begin{equation}
|i\rangle\left\langle\left. j\right|_{A} \otimes \mid k\right\rangle\left\langle\left. l\right|_{B} \otimes \mathbb{I}_{\text {dim }_{C}}.\right.
\end{equation}
The Kraus operator is
\begin{equation}
\begin{array}{l}
	K_{Z}=\left[\left(\begin{array}{l}
		1 \\
		0
	\end{array}\right) \otimes\left(\begin{array}{llll}
		1 & & & \\
		& 0 & & \\
		& & 0 & \\
		& & & 0
	\end{array}\right)+\left(\begin{array}{l}
		0 \\
		1
	\end{array}\right) \otimes\left(\begin{array}{llll}
		0 & & & \\
		& 1 & & \\
		& & 0 & \\
		& & & 0
	\end{array}\right)\right] \\
	\otimes\left(\begin{array}{llll}
		1 & & & \\
		& 1 & & \\
		& & 0 & \\
		& & & 0
	\end{array}\right) \otimes\left(\begin{array}{lll}
		1 & & \\
		& 1 & \\
		& & 0
	\end{array}\right) \otimes\left(\begin{array}{l}
		1 \\
		0
	\end{array}\right), \\
	K_{X}=\left[\left(\begin{array}{l}
		1 \\
		0
	\end{array}\right) \otimes\left(\begin{array}{llll}
		0 & & & \\
		& 0 & & \\
		& & 1 & \\
		& & & 0
	\end{array}\right)+\left(\begin{array}{l}
		0 \\
		1
	\end{array}\right) \otimes\left(\begin{array}{llll}
		0 & & & \\
		& 0 & & \\
		& & 0 & \\
		& & & 1
	\end{array}\right)\right] \\
	\otimes\left(\begin{array}{llll}
		0 & & & \\
		& 0 & & \\
		& & 1 & \\
		& & & 1
	\end{array}\right) \otimes\left(\begin{array}{lll}
		1 & & \\
		& 1 & \\
		& & 0
	\end{array}\right) \otimes\left(\begin{array}{l}
		0 \\
		1
	\end{array}\right). \\
\end{array}
\end{equation}
while the key maps are
\begin{equation}
\begin{array}{l}
	Z_{1}=\left(\begin{array}{ll}
		1 & 0 \\
		0 & 0
	\end{array}\right) \otimes \mathbb{I}_{\operatorname{dim}_{A} \times \operatorname{dim}_{B} \times \operatorname{dim}_{C} \times 2}, \\
	Z_{2}=\left(\begin{array}{ll}
		0 & 0 \\
		0 & 1
	\end{array}\right) \otimes \mathbb{I}_{\operatorname{dim}_{A} \times \operatorname{dim}_{B} \times \operatorname{dim}_{C} \times 2}.
\end{array}	
\end{equation}
\section{Channel model}\label{Channelmodel} 
 In this section, we will provide a description of the channel model used in our simulation. The channels utilized in our simulation include loss, misalignment, and dark count rate channels, which are similar to ~\cite{2021Wangwenyuan}. \blk
\subsection{BB84}\label{BB84channel}
In the WCP source, the output is a coherent state with amplitude of $\mu$. After passing through the misaligned channel, the amplitude reaching each detector can be summarized as
\begin{equation}
\begin{array}{|c|c|cccc|}
	\hline & & \multicolumn{4}{|c|}{\text { Bob's detectors (passive detection) }} \\
	\hline & & \mathrm{Z_{+}} & \mathrm{Z_{-}} &  \mathrm{X_{+}} &  \mathrm{X_{-}} \\
	\hline & \mathrm{Z_{+}} & \sqrt{p_{Z}} \cos \theta & \sqrt{p_{Z}} \sin \theta & \sqrt{p_{X}} \cos \alpha & \sqrt{p_{X}} \sin \alpha \\
   \text {Alice}	& \mathrm{Z_{-}}  & -\sqrt{p_{Z}} \sin \theta & \sqrt{p_{Z}} \cos \theta & \sqrt{p_{X}} \sin \alpha & -\sqrt{p_{X}} \cos \alpha \\
	\text {sends} &\mathrm{X_{+}} & \sqrt{p_{Z}} \sin \alpha & \sqrt{p_{Z}} \cos \alpha & \sqrt{p_{X}} \cos \theta & -\sqrt{p_{X}} \sin \theta \\
	& \mathrm{X_{-}} & \sqrt{p_{Z}} \cos \alpha & -\sqrt{p_{Z}} \sin \alpha & \sqrt{p_{X}} \sin \theta & \sqrt{p_{X}} \cos \theta \\
	\hline
\end{array}.
\end{equation}
Here, $\alpha =\frac{\pi}{4} -\theta $, $\theta$ is the  misalignment. Considering the channel loss, the loss factor $\sqrt{\mu\eta } $ should be multiplied before the above amplitudes. By considering the dark count, we can get the click probability of each detector:

\begin{equation}
	p_{j \mid i}^{\text {click }}=1-\left(1-p_{d}\right) \times e^{-\left|\alpha_{j|i}\right|^{2}},
\end{equation}
where $\alpha_{j|i}$ is the amplitude reaching the detector, $p_{d}$ is the detector dark count rate $i,j\in \left \{  H,V,+,-\right \}$.  The probabilities of individual detector clicks are known, for a given $i$, there could be a total of 4 detectors that will register a click, leading to 16 possible detection patterns. The probability of each detection pattern $ b_{1} b_{2} b_{3} b_{4} $ is represented by \blk
\begin{equation}
p_{b_{1} b_{2} b_{3} b_{4} \mid i}=\Pi_{j=1,2,3,4}\left\{\overline{b_{j}}+p_{j \mid i}^{\text {click }}(-1)^{\overline{b_{j}}}\right\},	
\end{equation}
where $b_{k} $ represents the response of the $k$ detectors, $b_{k}=0,1 $. $\overline{b_{k}} $ is the bit flip of $\overline{b_{k}}$.

For a given signal intensities $\mu$ ($\mu \in\left \{ u, v, w \right \}  $), iterate through all the $i$ and all of the detection mode to obtain $4\times 16$ data, and then write it into a matrix $P_{raw,\mu}$ with $4\times 16$ data.

Suppose that double click events on the same basis are randomly assigned to a measurement value, while double click events on different basis  are discarded. The following deletion model is defined:
\begin{equation}
	\begin{aligned}
		M_{H} &=[0,0,0,0,0,0,0,0,1,0,0,0,0.5,0,0,0] \\
		M_{V} &=[0,0,0,0,1,0,0,0,0,0,0,0,0.5,0,0,0] \\
		M_{+} &=[0,0,1,0.5,0,0,0,0,0,0,0,0,0,0,0,0] \\
		M_{-} &=[0,1,0,0.5,0,0,0,0,0,0,0,0,0,0,0,0] \\
		M_{\varnothing} &=[1,0,0,0,0,1,1,1,0,1,1,1,0,1,1,1] \\
		M &=\left[M_{H}^{T}, M_{V}^{T}, M_{+}^{T}, M_{-}^{T}, M_{\varnothing}^{T}\right].
	\end{aligned}
\end{equation}
Then the simulated statistical data can be given by
\begin{equation}
	P_{\mu}=P_{\text {raw }, \mu} \times M.
\end{equation}
\subsection{MDI}\label{MDIchannel}
In the case of MDI, the WCP source of Alice and Bob transmits a weak coherent state with amplitude of $\mu$. After passing through the misaligned channel, the signals of Alice and Bob are mismatched
$\theta _{A}$, $\theta _{B}$. Because $H$ and $V$ are different modes, we can simply think of $\sqrt{\mu_{A}} \cos \theta _{A} $ in H mode (similar to Alice and Bob), $\sqrt{\mu_{A}} \sin \theta _{A} $ in mode V(Alice and Bob are similar), then the amplitude reaching each detector is expressed as follows:
\begin{equation}
	\begin{aligned}
		\alpha_{3 H}^{\phi} &=\sqrt{\mu_{A} \eta_{A} \cos \theta_{A} / 2}+i \sqrt{\mu_{B} \eta_{B} \cos \theta_{B} / 2} e^{i \phi} \\
		\alpha_{4 H}^{\phi} &=i \sqrt{\mu_{A} \eta_{A} \cos \theta_{A} / 2}+\sqrt{\mu_{B} \eta_{B} \cos \theta_{B} / 2} e^{i \phi} \\
		\alpha_{3 V}^{\phi} &=\sqrt{\mu_{A} \eta_{A} \sin \theta_{A} / 2}+i \sqrt{\mu_{B} \eta_{B} \sin \theta_{B} / 2} e^{i \phi} \\
		\alpha_{4 V}^{\phi} &=i \sqrt{\mu_{A} \eta_{A} \sin \theta_{A} / 2}+\sqrt{\mu_{B} \eta_{B} \sin \theta_{B} / 2} e^{i \phi}.
	\end{aligned}
\end{equation}
Then the click probability of each detector can be given by
\begin{equation}
p_{k \mid i j}^{\text {click }, \phi}=1-\left(1-p_{d}\right) \times e^{-\left|\alpha_{k \mid i j}^{\phi}\right|^{2}},
\end{equation}
where $\alpha_{k \mid i j}^{\phi}$ is the amplitude reaching the detector, $i,j\in \left \{  H,V,+,-\right \}$, $k\in \left \{  3H,3V,4H,4V\right \}$.

For fixed $i,j$, a total of 4 detectors may respond, which results in a total of 16 possible detection modes. The response probability of each detection mode $ b_{1} b_{2} b_{3} b_{4} $ is given by

\begin{equation}
p_{b_{1} b_{2} b_{3} b_{4} \mid i j}=\prod_{k=1,2,3,4}\left\{\overline{b_{k}}+p_{k \mid i j}^{\text {click }}(-1)^{\overline{b_{k}}}\right\},
\end{equation}
where $b_{k} $ represents the response of the $k$ detectors, $b_{k}=0,1 $. $\overline{b_{k}} $ is the bit flip of $\overline{b_{k}}$.

For a given signal intensities $\mu_{A}\mu_{B}$ ($\mu_{A},\mu_{B} \in\left \{ u, v, w \right \}  $), Traverse all $i,j$ and detection modes, there are $ 4\times 4\times 16$ data in total. Write the data as  $P_{raw,\mu_{A}\mu_{B}}$ and define the following deletion model
\begin{equation}
\begin{aligned}
	M_{\Psi^{-}} &=[0,0,0,0,0,0,1,0,0,1,0,0,0,0,0,0] \\
	M_{\Psi^{+}} &=[0,0,0,1,0,0,0,0,0,0,0,0,1,0,0,0] \\
	M_{\varnothing} &=1_{1 \times 16}-M_{\Psi^{+}}-M_{\Psi^{-}} \\
	M &=\left[M_{\Psi^{-}}^{T}, M_{\Psi^{+}}^{T}, M_{\varnothing}^{T}\right].
\end{aligned}
\end{equation}
Then the simulated statistical data can be given by
\begin{equation}
P_{\mu_{A} \mu_{B}}=P_{\text {raw, } \mu_{A} \mu_{B}} \times M.
\end{equation}

\end{document}